\documentclass[mypaper,8pt,twoside]{CoAst}
\usepackage{epsf,graphicx,fancyhdr,sfmath}
\renewcommand{\chaptermark}[1]{\markboth{#1}{}}

\newcommand{\logg}{\ensuremath{\log g}}                     

\newcommand{\kopf}{\small\itshape Comm.\ in Asteroseismology, N$^{\textsf{\underline{o}}}$ 159, 2009\\
Proceedings of the JENAM 2008 Symposium N$^{\textsf{\underline{o}}}$~4:
Asteroseismology and Stellar Evolution}

\newcommand{\Authors}[1]{\begin{center}\normalsize\bf\sf #1 \end{center}}

\renewcommand{\author}[1]{\begin{center}\normalsize\bf\sf #1 \end{center}}
\newcommand{\Address}[1]{\begin{center}\small\sf #1 \end{center}}

\newcommand{\Objects}[1]{{\vspace{3mm}\small \noindent Individual Objects: }\small\sf \hangindent=27truemm \hangafter=1 #1 }

\renewenvironment{abstract}{\section*{Abstract}\normalsize\sf}{}
\newcommand{\References}[1]{\begin{flushleft}{\large References\\}\vspace*{2mm}\small #1 \end{flushleft}}

\newcommand{\chapterCoAst}[2]{\chapter[\sf\normalsize #1\\ \footnotesize \hspace*{5mm}by #2 \sf\normalsize][]{#1\\}\rhead[\fancyplain{}{\sf\footnotesize \center{#1}}]{\fancyplain{}{\sffamily\thepage}}\lhead[\fancyplain{\kopf}{\sffamily\thepage}]{\fancyplain{\kopf}{\sf\footnotesize \center{#2}}}}

\renewcommand{\chaptername}{}

\newcommand{\acknowledgments}[1]{\vspace*{5mm}\noindent  \textbf{Acknowledgments.} #1}

\def\rfr{\smallskip\par\noindent
        \hangindent=7truemm
        \hangafter=1}

\begin{document}
\sf

\chapterCoAst{Long-term EXOTIME photometry and follow-up spectroscopy of the sdB
  pulsator HS\,0702+6043}
{R.\,Lutz, S.\,Schuh, R.\,Silvotti, et al.} 
\Authors{R.\,Lutz$^{1,2}$, S.\,Schuh$^{1}$, R.\,Silvotti$^3$, R.\,Kruspe$^1$,
  and S.\,Dreizler$^1$} 
\Address{
$^1$ Institut f\"ur Astrophysik, Friedrich-Hund-Platz 1, 37077 G\"ottingen, Germany\\
$^2$ Max-Planck-Institut f\"ur Sonnensystemforschung, Max-Planck-Stra\ss e 2,\\
  37191 Katlenburg-Lindau, Germany\\
$^3$ INAF - Osservatorio Astronomico di Capodimonte, via Moiariello 16,\\80131
  Napoli, Italy
}

\noindent
\begin{abstract}
Pulsating subdwarf B (sdB) stars oscillate in short-period $p$-modes or
long-period $g$-modes. \mbox{HS\,0702+6043} (DW Lyn) is one of a few objects 
to show characteristics of both types and is hence classified as 
hybrid pulsator. It is one of our targets in the EXOTIME program 
to search for planetary companions around extreme horizontal branch 
objects. In addition to the standard exercise in asteroseismology to probe 
the instantaneous inner structure of a star, measured changes in the pulsation 
frequencies as derived from an O--C diagram can be compared to 
theoretical evolutionary timescales. Based on the photometric data available
so far, we are able to derive a high-resolution frequency spectrum and to 
report on our efforts to construct a multi-season O--C diagram. Additionally, 
we have gathered time-resolved spectroscopic data in order to constrain stellar 
parameters and to derive mode parameters as well as radial and 
rotational velocities.
\end{abstract}

\Objects{HS\,0702+6043, HS\,2201+2610, HW\,Vir}

\section*{Timing Method and the EXOTIME program}
The O--C analysis (Observed minus Calculated) is a tool to
measure the phase variations of a periodic function. The observed times of the
pulsation maxima of single runs are compared to the calculated mean ephemeris of the
whole data set. Since a low-mass companion, due to its gravitational
influence, would cause cyclically advanced and
delayed timings of an oscillating sdB's pulsation maxima (due to motion around the common
barycenter), this method can be used to search for exoplanets. A sinusoidal
component in an O--C diagram is therefore a signature for a companion. In
addition, this method can be used to derive evolutionary timescales by
measuring linear changes in the pulsation periods. The signature would in this
case be a parabolic shape in the O--C diagram. Using this timing method, Silvotti
et al.\ (2007) detected a giant planet companion to the pulsating subdwarf B
star HS\,2201+2610 (V\,391 Peg) and recently Lee et al.\ (2008) reported two
planets around the eclipsing sdB+M binary system HW\,Vir, also revealed by an
O--C diagram analysis by measuring the timings of the eclipse minima.\\
The EXOplanet search with the TIming MEthod (EXOTIME) program is an internationally
coordinated effort to examine pulsating sdB stars in terms of planetary
companions and evolutionary aspects. Closely related to the puzzling evolution
of sdB stars is the late-stage or post-RG evolution of planetary systems and
the question if planets could be responsible for the extreme mass loss of the
sdB progenitors (e.g.\ Soker 1998). EXOTIME performs ground-based time-series
photometry from various sites with telescopes in the 0.5\,m to 3.6\,m range.

\begin{table}[!thb]
\begin{center}
\sffamily
\caption{Our current photometric data archive for HS\,0702+6043. Sites: Calar
  Alto 2.2m/1.2m (CA2/1), NOT 2.56m (N), G\"ottingen 0.5m (G), T\"ubingen 0.8m (T), Steward
  Bok 2.2m (SB), Loiano 1.5m (L), Mt.Bigelow 1.55m (MB), Konkoly 1m (K).} 
\label{tab:dataarchive}
\medskip
\begin{tabular}{lcr|lcr}
\hline
Date               & Site  & Length & Date               & Site  & Length\\ 
                   &       & [h]    &                    &       & [h]   \\
\hline
December 1999     & CA1 & 8.4   & Nov07-Mar08  & MB  & 424.0\\
October 2000      & N   & 0.7   & March 2008 & G  & 0.6 \\
February 2004     & T   & 7.3   & March 2008 & L  & 8.0 \\
February 2004     & SB  & 12.0  & March 2008 & K  & 11.5\\
January 2005     & CA2 & 56.0  & May 2008 & G  & 4.1 \\
December 2007     & T   & 31.8  & September 2008 & G  & 3.3 \\
December 2007     & G   & 12.0  & October 2008 & K  & 2.6 \\
February 2008     & T   & 20.1  & October 2008 & CA2& 4.2 \\
February 2008     & G   & 32.2  & November 2008 & CA2& 8.7 \\
\hline
\end{tabular}
\end{center}
\end{table}

\section*{Long-term Photometry}
HS\,0702+6043 was first identified as a variable by Dreizler et al.\ (2002).
It is placed at the common boundary of the $p$- and $g$-mode instability
regions in a $\logg$-$T_{\textrm{eff}}$ diagram. The two strongest $p$-mode 
pulsations at 363.11\,s and 383.73\,s (amplitudes of 30 and 6\,mmag,
respectively) will be used to construct multi-seasonal O--C diagrams, for
which a timebase of several years is needed. For deriving a single O--C point, 
at least three to four consecutive nights of observations are needed to
provide a sufficient frequency resolution. We aim for a minimum of six O--C 
points per year. Our data archive for HS\,0702+6043 dates back to 1999, 
unfortunately with large gaps in between. Table 1 lists our current
photometric data archive, not yet sufficient to present a meaningful O--C 
diagram due to the large gap between the observations in 1999 and 2004. 

\section*{Follow-up Spectroscopy}
The 772 time resolved high-resolution Echelle spectra (20\,s each) of HS\,0702+6043
taken at the Hobby Eberly Telescope will provide rotational and in particular
pulsational radial velocities. The pulsational amplitudes are expected to be
larger for HS\,0702+6043 than for HS\,2201+2610 since the photometric
amplitudes are larger. An analysis of the time-resolved spectroscopy of HS\,2201+2610 can
be found in Schuh et al.\ (2008).

\acknowledgments{
  The authors thank all observers who contributed observations to the 
  HS\,0702+6043 data archive: B.Beeck, Z.Bognar, E.M.Green and collaborators, 
  M.Hundertmark, T.Nagel, R.\O stensen, M.Paparo, P.Papics, T.Stahn.
  Partly based on observations collected in service
  mode by L.Montoya, M.Alises and U.Thiele for our program
  \mbox{H08-2.2-009} at the Centro Astron{\'o}mico Hispano
  Alem{\'a}n (CAHA) at Calar Alto, operated jointly by the Max-Planck
  Institut f\"ur Astronomie and the Instituto de Astrof{\'i}sica de
  Andaluc{\'i}a (CSIC).
  The authors also thank the Astronomische Gesellschaft as well as the
  conference sponsors and in particular HELAS
  (European Helio- and Asteroseismology Network, an European
  initiative funded by the European Commission since April 1st, 2006,
  as a ''Co-ordination Action'' under its Sixth Framework Programme FP6)
  for financially supporting the poster presentation at JENAM 2008
  Minisymposium N$^{\circ}$~4 through travel grants to R.L.\ and S.S. 
}

\References{
\rfr Dreizler, S., Schuh, S., Deetjen, J.L., et al.\ 2002, A\&A, 386, 249
\rfr Lee, J.~W., Kim, S.-L., Kim, C.-H., et al.\ 2008, arXiv, 0811.3807
\rfr Schuh, S., Kruspe, R., Lutz, R., \& Silvotti, R.\ 2008, these proceedings
\rfr Silvotti, R., Schuh, S., Janulis, R., et al.\ 2007, Nature, 449, 189
\rfr Soker, N.\ 1998, AJ, 116, 1308
}

\end{document}